\def\draftversion{false}

\RequirePackage{ifthen}
\ifthenelse{\equal{\draftversion}{true}}{
  \documentclass[aps,prb,10pt,galley,amsmath,amssymb,showpacs,
                 superscriptaddress]{revtex4}
}{
  \documentclass[aps,prb,10pt,twocolumn,amsmath,amssymb,showpacs,
                 superscriptaddress]{revtex4-1}
}

\usepackage{graphicx}
\usepackage{xfrac}
\usepackage{bm} 
\usepackage{bbm}

\def\nn{\nonumber\\}
\def\beq{\begin{equation}}
\def\eeq{\end{equation}}
\def\bea{\begin{eqnarray}}
\def\eea{\end{eqnarray}}

\def\eig{\epsilon}

\def\uu{^{(\rm u)}}
\def\occ{^{\rm occ}}
\def\unf{^{\rm unf}}
\def\part{\wt{\partial}}

\def\Re{{\rm Re\,}}
\def\Im{{\rm Im\,}}
\def\tr{{\rm tr\,}}
\def\Tr{{\rm Tr\,}}

\def\PP{\mathbbm{P}}
\def\QQ{\mathbbm{Q}}
\def\FF{\mathbbm{F}}
\def\T{T}

\def\a{{\bf a}}
\def\A{{\bf A}}

\def\k{{\bf k}}
\def\K{{\bf K}}
\def\G{{\bf G}}
\def\rr{{\bf r}}
\def\R{{\bf R}}
\def\M{{\bf M}}
\def\O{{\bf 0}}
\def\op{{\cal O}}

\def\ket#1{\vert#1\rangle}
\def\bra#1{\langle#1\vert}
\def\ip#1#2{\langle#1\vert#2\rangle}
\def\me#1#2#3{\langle#1\vert#2\vert#3\rangle}

\def\wt#1{\widetilde{#1}}

\def\im{\mathrm{Im}\,}

\def\HH{\mathbbm{H}}
\def\AAA{\mathbbm{A}}

\newcommand{\eq}[1]{Eq.~(\ref{eq:#1})}
\newcommand{\eqs}[2]{Eqs.~(\ref{eq:#1}) and~(\ref{eq:#2})}
\newcommand{\equ}[1]{Equation~(\ref{eq:#1})}

\begin{document}

\title{How disorder affects the Berry-phase anomalous Hall
  conductivity: the view from $k$~space}

\author{Raffaello Bianco}
\affiliation{
Dipartimento di Fisica, Universit\`a di Trieste, 34127 Trieste, Italy}
\affiliation{Centro de F\'{\i}sica de Materiales, 
Universidad del Pa\'{\i}s Vasco, 20018 San Sebasti\'an, Spain}
\author{Raffaele Resta}
\affiliation{
Dipartimento di Fisica, Universit\`a di Trieste, 34127 Trieste, Italy}
\affiliation{
Donostia International Physics Center, 20018 San Sebasti\'an, Spain}
\author{Ivo Souza}
\affiliation{Centro de F\'{\i}sica de Materiales, 
Universidad del Pa\'{\i}s Vasco, 20018 San Sebasti\'an, Spain}
\affiliation{Ikerbasque Foundation, 48011 Bilbao, Spain}



\date{\today}
\begin{abstract}
  The anomalous Hall conductivity of ``dirty'' ferromagnetic metals is
  dominated by a Berry-phase contribution which is usually interpreted
  as an intrinsic property of the Bloch electrons in the pristine
  crystal. In this work we evaluate the geometric Hall current
  directly from the electronic ground state with disorder, and then
  recast it as an integral over the crystalline Brillouin zone.  The
  integrand is a generalized $k$-space Berry curvature, obtained by
  unfolding the Berry curvature from the small Brillouin zone of a
  large supercell.  Therein, disorder yields a net extrinsic Hall
  contribution, which we argue is related to the elusive side-jump
  effect.  As an example, we unfold the first-principles Berry
  curvature of an Fe$_3$Co ordered alloy from the original fcc-lattice
  Brillouin zone onto a bcc-lattice zone with four times the
  volume. Comparison with the virtual-crystal Berry curvature clearly
  reveals the symmetry-breaking effects of the substitutional Co
  atoms.

\end{abstract}

\maketitle


\section{Introduction}
\label{sec:intro}

The anomalous Hall effect (AHE) in metallic ferromagnets includes a
purely geometric bandstructure contribution given by the $k$-space
Berry curvature of the occupied Bloch
states.\cite{nagaosa-rmp10,xiao-rmp10} Somewhat counterintuitively,
this {\it intrinsic} contribution only becomes dominant in moderately
resistive (``dirty'') samples, for which crystal momentum is not a
good quantum number and the Berry curvature strictly speaking is
ill-defined.  In highly-conducting pristine samples with sharply
defined energy bands and Berry curvature, the AHE is instead dominated
by an {\it extrinsic} contribution, skew-scattering from dilute
impurities.\cite{nagaosa-rmp10,xiao-rmp10}

Motivated by these considerations, we introduce a generalized
$k$-space Berry curvature for metallic systems with disorder.
Integrated over the Brillouin zone (BZ) of the underlying ordered cell
it gives the dominant contribution to the anomalous Hall conductivity
(AHC), expressed as a property of the disordered electronic ground
state. We will call this the {\it geometric} AHC contribution. It
combines the nominally intrinsic contribution with certain disorder
effects of a similar nature.  With this definition we depart from the
standard terminology, where the words ``intrinsic'' and ``geometric''
(or ``Berry-phase'') are used interchangeably when refering to AHC
contributions.\cite{nagaosa-rmp10,xiao-rmp10} The proposed definition
has the merit of being directly applicably to the experimental regime
of interest, where scattering from disorder is important.

Our generalized Berry-curvature definition is based on the notion of
BZ unfolding, which has been used extensively in recent years in the
context of band structure calculations with periodic
supercells.\cite{ku-prl10,popescu-prb12,allen-prb13} Similar unfolding
techniques were introduced long ago to describe the phonon spectra of
disordered alloys.\cite{baroni-prl90}

Unfolding band structures of supercell (SC) calculations is a
particularly informative way of visualizing the influence of
impurities (or other sources of broken translational order) on the
electronic states in crystals. For weak to moderate disorder the
unfolded bands resemble those of the pristine crystal, with the
deviations in both the dispersion and the spectral weight reflecting
the effect of the disorder potential.\cite{ku-prl10} The recent
development of efficient {\it ab-initio}-based SC
methodologies\cite{berlijn-prl11} opens up new possibilities for
applying unfolding techniques to large SCs with realistic descriptions
of disorder.\cite{berlijn-prl12,liu-prb13}

So far, BZ unfolding has been used mainly to extract approximate
energy dispersions for disordered systems. While the energy bands
$\eig_{\k i}$ are the most basic quantity in the theory of solids, it
is now understood that the $k$-space Berry curvature
$\boldsymbol{\Omega}_i(\k)=\boldsymbol{\nabla}_\k\times \A_i(\k)$ is
an additional fundamental ingredient determining the dynamics of
electrons in crystals.\cite{xiao-rmp10} (Here $\A_i(\k)$ is the Berry
connection, to be defined shortly.)  Using Stokes' theorem, the Berry
curvature can be viewed as the geometric phase $\varphi_i=\oint
\A_i(\k)\cdot d{\bf l}$ per unit area picked up by a Bloch electron in
band $i$ as it is transported adiabatically along a small loop in
$k$-space.  The Berry curvature is generically nonzero in the BZ of
crystals with broken inversion or time-reversal symmetry. It modifies
the motion of electron wavepackets driven by an electric field
$\bm{E}$, by adding a transverse "anomalous velocity" term
$(e/\hbar)\boldsymbol{\Omega}_i(\k)\times \bm{E}$ to the usual band
velocity $(1/\hbar)\boldsymbol{\nabla}_\k \epsilon_{\k j}$.

The intrinsic AHC is a direct consequence of the anomalous
velocity. It is given by\cite{xiao-rmp10,nagaosa-rmp10}
\beq
\label{eq:ahc}
\sigma^{\rm int}_{ab}=-\frac{e^2}{h}\int_{\rm NBZ}
\frac{d^3k}{(2\pi)^3}\Omega\occ_{ab}(\k)
\eeq
\beq
\label{eq:curv-tot}
\Omega\occ_{ab}(\k)=\sum_i
f_{\k i}\Omega_{i,ab}(\k)
\eeq
\beq
\label{eq:curv-band}
\Omega_{i,ab}(\k)=\epsilon_{abc}\Omega_{i,c}(\k)
=\nabla_{k_a} A_{i,b}(\k)-\nabla_{k_b} A_{i,a}(\k)\,,
\eeq
where $f_{\k i}$ is the occupation of the Bloch eigenstate $\ket{\k
  i}=e^{i\k\cdot\hat{\bf x}}\ket{u_{\k i}}$ 
and $\A_i(\k)=i\ip{u_{\k i}}{\boldsymbol{\nabla}_\k {u_{\k i}}}$ is
the Berry connection of the $i$-th band.  The integral in \eq{ahc} is
over the BZ of the pristine crystal, which we will call the ``normal
Brillouin zone'' (NBZ).

The definition of the intrinsic AHC as a Berry curvature in $k$ space
relies on perfect translational order.  This is at odds with the
above-mentioned fact that the intrinsic contribution tends to dominate
in dirty samples with broken translational invariance. The
conventional formulation becomes even more problematic for
intrinsically disordered systems such as random alloys, for which
there is no experimentally accessible ``clean limit.''  And yet, it is
still useful to reason in terms of ``intrinsic'' contributions to the
AHE in such moderately conducting systems.\cite{lowitzer-prl10}

In view of these difficulties, how should the intrinsic AHC be defined
and calculated in the presence of disorder? The standard procedure is
to define it in terms of the Berry curvature of an ordered reference
system --~the pristine crystal in the case of doped
samples,\cite{fang-science03,yao-prb07} or a ``virtual crystal''
effective Hamiltonian in the case of alloys\cite{jungwirth-prl02}~--
calculated using the band filling appropriate to the doping level or
alloying concentration.  Disorder effects can be included via a
diagonal self-energy term inserted in the energy denominator of the
sum-over-states expression for the Berry curvature [\eq{curv-kubo-tot}
below], to account for the finite lifetime of the Bloch
eigenstates.\cite{yao-prb07} A related strategy, which has been
implemented within the coherent-potential approximation, is to compute
the intrinsic AHC starting from the Kubo-Str\u{e}da equation, by
combining all terms not connected to vertex
corrections.\cite{lowitzer-prl10} While physically motivated, these
remain somewhat {\it ad-hoc} and model-dependent prescriptions, which
can only be justified for sufficiently dilute or concentrated alloys.

We propose a different approach, where we do not insist on defining
precisely the intrinsic AHC contribution in a disordered system, and
replace it with the geometric AHC.  In the SC approach it is computed
by inserting into \eq{ahc} the electronic states of the SC system,
\beq
\label{eq:ahc-SBZ}
\sigma^{\rm geom}_{ab}=-\frac{e^2}{h}\int_{\rm SBZ}
\frac{d^3K}{(2\pi)^3}\Omega\occ_{ab}(\K)\,,
\eeq
and averaging over several realizations of disorder. Here
$\Omega\occ_{ab}(\K)$ is the Berry curvature of the occupied SC
eigenstates $\ket{\K J}$, and the integral is over the supercell
Brillouin zone (SBZ). No phenomenological lifetime broadening
parameter needs to be included in the Berry curvature calculation,
since spectral broadening by disorder is already built-in, as revealed
by the configuration-averaged unfolded energy
bands.\cite{berlijn-prl11,berlijn-prl12,liu-prb13}

\equ{ahc-SBZ} is a very plausible generalization of \eq{ahc} in the
context of SC calculations, where a disordered system is modeled as a
``crystal'' with a very large ``primitive cell.'' It correctly gives a
quantized value for the AHC of 2D disordered Chern insulators when the
Fermi level lies in the mobility gap,\cite{prodan-prl10,zhang-cpb13}
and we propose to use it to unambiguously identify a dominant
contribution to the AHC of {\it metallic} disordered systems.
(Contrary to the case of Chern insulators, \eq{ahc-SBZ} does not
capture the full AHC of a metal in a finite SC; we will return to this
point in Sec.~\ref{sec:disc}.)

Realistic descriptions of disorder require reasonably large SCs. The
integration volume in \eq{ahc-SBZ} then becomes very small, and all
$k$-space information is lost.  In order to restore a 
$k$-space description reminiscent of \eq{ahc}, we recast \eq{ahc-SBZ}
as the NBZ integral of a suitably defined ``unfolded Berry
curvature,''
\beq
\label{eq:ahc-NBZ}
\sigma^{\rm geom}_{ab}=-\frac{e^2}{h}\int_{\rm NBZ}
\frac{d^3k}{(2\pi)^3}\Omega\unf_{ab}(\k)\,.
\eeq
Although with disorder present the unfolded curvature is no longer
geometric in the strict sense (the interpretation as a Berry phase per
unit area is lost), it remains gauge invariant in the NBZ. To
illustrate its behavior in metallic systems with reduced translational
order, we will implement \eq{ahc-NBZ} from first-principles, and apply
it to a simple test case of an ordered magnetic alloy.

The manuscript is organized as follows. In Sec.~\ref{sec:spectral} we
motivate our approach starting from the Berry curvature defined in the
folded BZ of a disordered SC. In Sec.~\ref{sec:formalism} we introduce
a general BZ unfolding formalism, which we then apply to the Berry
curvature; the details of the implementation in a Wannier-function
basis are also given. In Section~\ref{sec:results} we compute from
first-principles the unfolded curvature of an ordered Fe$_3$Co alloy,
and compare it with the Berry curvatures of pure bcc Fe and of the
alloy in the virtual-crystal approximation (VCA). We conclude in
Section~\ref{sec:disc} with a discussion and an outlook.

\section{Berry curvature in the folded Brillouin zone}
\label{sec:spectral}

The formal connection between the geometric and linear-response
formulations of the intrinsic AHC is provided by the spectral
representation of \eq{curv-tot},
\beq
\label{eq:curv-kubo-tot}
\Omega\occ_{ab}(\k)=-\im\sum_{i,j}\,(f_{\k i}-f_{\k j})
\frac{\me{\k i}{\hbar\hat{v}_a}{\k j}
      \me{\k j}{\hbar\hat{v}_b}{\k i}}
     {(\epsilon_{\k j}-\epsilon_{\k i})^2}\,,
\eeq
whose NBZ integral (\ref{eq:ahc}) is equivalent to the Kubo-Greenwood
formula for the AHC in the clean limit.\cite{nagaosa-rmp10}

\equ{curv-kubo-tot} is written in terms of the Bloch eigenstates and
energy eigenvalues of a pristine crystal. If we place the crystal in a
periodic SC and introduce some 
disorder, the disorder potential mixes states with different NBZ
momenta $\k$ and $\k'$ whenever $\k'-\k$ equals a SC reciprocal vector
$\G$, forcing the new eigenstates to be labeled by a common
wavevector $\K$ in the SBZ.  The Berry curvature can still be defined
in the SBZ from \eq{curv-kubo-tot}, now written in terms of the SC
eigenstates $\ket{\K J}$, energy eigenvalues $\epsilon_{\K J}$, and
occupations $f_{\K J}$.

Because of those extra couplings from disorder, it is not obvious how
to map (unfold) the Berry curvature from the SBZ onto the NBZ of the
original crystal. Clearly, $\boldsymbol{\Omega}\occ(\K)$ is not simply
equal to the virtual-crystal Berry curvature summed over the points
$\{\k_s\}$ which fold onto~$\K$: $\boldsymbol{\Omega}\occ(\K)\not=
\sum_s\,\boldsymbol{\Omega}\occ_{\rm VCA}(\k_s)$.  Nevertheless, it
will be possible to arrive at a unique definition for the unfolded
Berry curvature with all the desired properties, namely: (i) it
reduces to the ordinary Berry curvature $\boldsymbol{\Omega}\occ(\k)$
in the clean limit; (ii) it remains sharply defined (gauge-invariant)
in the presence of disorder; and (iii) it constitutes a proper mapping
from the SBZ to the NBZ in the sense that
\beq
\label{eq:curv-sum-rule}
\boldsymbol{\Omega}\occ(\K)=\sum_s\,\boldsymbol{\Omega}\unf(\k_s)\,,
\eeq
which provides the link between \eqs{ahc-SBZ}{ahc-NBZ}.  

The difference between the unfolded SC curvature and the Berry
curvature of the virtual crystal with averaged disorder reflects the
disorder-mediated couplings between the folded bands, made possible by
the relaxed crystal-momentum selection rule inside the SC.  Such
``pseudodirect'' transitions\cite{popescu-prb12} modify the interband
coherence effects described by \eq{curv-kubo-tot}, giving additional
contributions to the anomalous velociy and AHC which are absorbed into
the definition of $\boldsymbol{\Omega}\unf(\k)$.

A striking feature of the Berry curvature in crystalline ferromagnets
is the occurence of sharp peaks when two energy bands lying on either
side of the Fermi level become
quasi-degenerate.\cite{fang-science03,yao-prl04,wang-prb06} This can
be understood in terms of \eq{curv-kubo-tot} as a resonant enhancement
behavior, and we will see that the same intuitive picture holds for
the unfolded quantities: strong peaks in $\boldsymbol{\Omega}\unf(\k)$
can be traced back to pairs of unfolded bands separated by small
(pseudo)direct gaps across $\epsilon_F$.

\section{Brillouin-zone unfolding}
\label{sec:formalism}

\subsection{Basic definitions}
\label{sec:basic}

Given a set of primitive translations $\{\a_i\}$ of the normal crystal
cell (NC), the SC primitive translations can be written as
$\sum_j\,M_{ij}\a_j$, with $\M$ an integer matrix.  Each point $\K$ in
the SBZ unfolds onto $|\M|$ distinct points $\k_s=\K+\G_s$ in the NBZ,
where the $\G_s$ are SC reciprocal lattice
vectors.\cite{popescu-prb12}

Following Ref.~\onlinecite{ku-prl10} we introduce a Bloch basis in the
NBZ, and another in the SBZ.  The basis states $\ket{\k n}$ and
$\ket{\K N}$ are normalized over the NC and the SC respectively, and
we define $\langle\ldots\rangle$ as an integral over the SC volume, so
that $\ip{\k n}{\k m}=|\M|\delta_{n,m}$.

We also define the projection operator
\beq
\label{eq:proj}
\hat{\T}(\k)=\frac{1}{|\M|}\sum_n\,\ket{\k n}\bra{\k n}\,.
\eeq
For any SC Bloch state we have
%
$\sum_s\hat{\T}(\k_s)\ket{\K N}=\ket{\K N}$,
%
which simply means that the state $\ket{\K N}$ has unfolded Bloch
character distributed among the points $\{\k_s\}$, with weights
$\me{\K N}{\hat{\T}(\k_s)}{\K N}$ which add up to
one.\cite{allen-prb13}

\subsection{Unfolding a generic $k$-space quantity}
\label{sec:unf-general}

Suppose we are interested in some property of the SC system which can
be calculated in the SBZ as the trace of a Hermitean matrix
\beq
\label{eq:op}
\op_{NM}(\K)=\me{\K N}{\hat{\op}}{\K M}\,.
\eeq
In order to map $\Tr \op(\K)$ from the SBZ onto the NBZ we first we
set up the matrix elements of $\hat{\op}$ in the Bloch basis at the
unfolded points,
\bea
\label{eq:mat-unfold}
\op\uu_{nm}(\k_s)&=&\frac{1}{|\M|}\me{\k_s n}{\hat{\op}}{\k_s m}\nn
&=&\frac{1}{|\M|}\sum_{N,M}\,S_{nN}(\k_s,\K)\op_{NM}(\K) \left[
  S^\dagger(\k_s,\K)\right]_{Mm},\nn
\eea
where $S_{nN}(\k_s,\K)=\ip{\k_s n}{\K N}$.  Let us also define
\bea
\label{eq:T}
\T_{MN}(\k_s,\K)&=&\me{\K M}{\hat{\T}(\k_s)}{\K N}\nn
&=&\frac{1}{|\M|}\left[ S^\dagger(\k_s,\K)S(\k_s,\K)\right]_{MN}\,,
\eea
whose diagonal elements are the unfolding weights.

The unfolded quantity is given by the trace of \eq{mat-unfold},
\beq
\label{eq:unfold}
\op\unf(\k_s)=
\tr {\cal O}\uu(\k_s)=\Tr\left[ \T(\k_s,\K)\op(\K)\right]\,,
\eeq
where ``tr'' and ``Tr'' denote traces over the NC and SC orbital
indices $n$ and $N$ respectively.  \equ{unfold} is our basic
prescription for BZ unfolding.  In Appendix~\ref{appendix:bands} we
verify that it correctly gives the unfolded energy bands.

\subsubsection{Gauge invariance of unfolded quantities}

Under a unitary mixing of the SC basis states,
\beq
\ket{\K N}\rightarrow\sum_M\,\ket{\K M}U_{MN}(\K)\,,
\eeq
the matrix (\ref{eq:T}) changes in a gauge-covariant manner,
\beq
\label{eq:gauge-cov}
\T(\k_s,\K)\rightarrow U^\dagger(\K)\T(\k_s,\K)U(\K)\,.
\eeq
If the matrix $\op(\K)$ is also gauge-covariant, then \eq{unfold}
remains unchanged under the transformation. This gauge-invariance
requirement will dictate which definition of a ``Berry curvature
matrix'' to use for unfolding purposes.  (While the matrix
representation (\ref{eq:op}) of most quantities is unique and
trivially gauge-covariant, the Berry curvature is more subtle, as it
involves $k$-space derivatives of the state vectors.)

\subsection{Unfolded Berry curvature}

Our goal is to unfold the Berry curvature of the SC system from the
SBZ to the NBZ.  Since the unfolding formalism developed above is
based on matrix objects, we begin by defining a Hermitean Berry
curvature matrix $\Omega_{ab,NM}(\K)=\Omega_{ab,MN}^*(\K)$ 
satisfying two essential requirement: (i) it should be gauge-covariant
in the sense of \eq{gauge-cov}, and (ii) its trace should give the
quantity to be unfolded: $\Tr\Omega_{ab}(\K) =\Omega\occ_{ab}(\K)$.

Those requirements are fulfilled by the non-Abelian Berry curvature
matrix.\cite{xiao-rmp10,ceresoli-prb06} For an insulator it reads
\beq
\label{eq:nonabelian-ins}
\Omega_{ab,NM}=\partial_a A_{b,NM}-\partial_b A_{a,NM}-i[A_a,A_b]_{NM}\,,
\eeq
where $\K$ has been dropped everywhere for brevity. Here
$\partial_a=\partial/\partial_{K_a}$, $A_{a,NM}=i\ip{u_N}{\partial_a
  u_M}$ is the Berry connection matrix, and the indices $N,M$ run over
the occupied states.  Except for the commutator, \eq{nonabelian-ins}
is the obvious matrix generalization of \eq{curv-band}. The extra term
does not affect the trace, but is needed to ensure gauge-covariance.

For our purposes it will be convenient to recast \eq{nonabelian-ins}
in terms of projection operators,\cite{ceresoli-prb06}
\beq
\label{eq:nonabelian}
\Omega_{ab,NM}=iF_{ab,NM}-iF_{ba,NM}\,,
\eeq
where
\beq
\label{eq:F-ab}
F_{ab,NM}=\me{u_N}{(\partial_a\hat{P})\hat{Q}(\partial_b\hat{P})}{u_M}=
F_{ba,MN}^*
\eeq
and $\hat{P}$, $\hat{Q}=\hat{\mathbbm{1}}-\hat{P}$ span the occupied
and unoccupied spaces respectively. Metals can be handled by writing
\beq
\label{eq:P}
\hat{P}=\sum_{N,M}\,\ket{u_N}f_{NM}\bra{u_M}\,,
\eeq
where $f_{NM}$ is the occupation matrix.\cite{lopez-prb12} For
insulators $\hat{P}=\sum_N^{\rm occ}\ket{u_N}\bra{u_N}$, and a few
lines of algebra show that \eq{nonabelian} reduces to
\eq{nonabelian-ins}.

With these definitions, the Berry curvatures in the original SBZ and
unfolded onto the NBZ via \eq{unfold} read
\beq
\label{eq:curv-sbz}
\Omega\occ_{ab}(\K)
=-2\Im\Tr F_{ab}(\K)
\eeq
\beq
\label{eq:curv-unf}
\Omega\unf_{ab}(\k_s)=-2\Im\Tr \left[T(\k_s,\K)F_{ab}(\K)\right]\,.
\eeq
\equ{curv-sbz} was given in Ref.~\onlinecite{ceresoli-prb06}, while
\eq{curv-unf} is a primary result of the present work.

It is easily verified (see Appendix~\ref{appendix:sum-rule}) that
\eq{curv-unf} satisfies the sum rule (\ref{eq:curv-sum-rule}), which
allows to recast the geometric AHC of the SC system as an integral
over the NBZ according to \eqs{ahc-SBZ}{ahc-NBZ}.

\subsection{Implementation in a Wannier basis}
\label{sec:wannier}

In this section we describe the implementation of \eq{curv-unf} using
Wannier interpolation, which is carried out as a post-processing step
following a first-principles SC calculation. Essentially, we combine
two Wannier-based methodologies: that of Refs.~\onlinecite{wang-prb06}
and~\onlinecite{lopez-prb12} for computing the Berry curvature, and
that of Ref.~\onlinecite{ku-prl10} for BZ unfolding.

In the formalism of Ref.~\onlinecite{ku-prl10} the Bloch basis
orbitals are chosen as $\ket{\K N}=\sum_\R\,e^{i\K\cdot\R}\ket{\R N}$,
where $\ket{\R N}$ is a Wannier function and $\R$ a SC lattice
vector. The Wannier functions are then mapped onto the NC according to
$\ket{\R N}\leftrightarrow \ket{\rr n}=\ket{\R+[\rr],n}$, with a
choice of $|\M|$ NC lattice vectors $[\rr]$ such that no two $[\rr]$'s
differ by an $\R$.  Once a map has been chosen, any NC lattice vector
$\rr$ can be uniquely decomposed as $\rr=\R+[\rr]$.  Setting $\ket{\k
  n}=\sum_\rr\,e^{i\k\cdot\rr}\ket{\rr n}$ then gives\cite{ku-prl10}
\beq
\label{eq:ovlp-wf}
S_{nN}(\k_s,\K)=\ip{\k_s n}{\K N}=e^{-i\k_s\cdot [\rr](N)}\delta_{n,n'(N)}\,,
\eeq
which goes into the unfolding equations (\ref{eq:T})
and~(\ref{eq:unfold}).

The expression for the unfolded Berry curvature involves several other
matrix objects, which we now define borrowing the notation from
Ref.~\onlinecite{lopez-prb12}. The two basic objects are (omitting
orbital indices)
\beq
\label{eq:HH-K}
\HH(\K)=\sum_\R\,e^{i\K\cdot\R}\me{\O}{\hat{H}}{\R}
\eeq
\beq
\label{eq:AA-K}
\AAA_a(\K)=\sum_\R\,e^{i\K\cdot\R}\me{\O}{\hat{x}_a}{\R}\,.
\eeq
Diagonalization of $\HH(\K)$ gives the energy eigenvalues,
\beq
\label{eq:HH-diag}
\HH^{({\rm H})}_{JJ'}(\K)=
\left[U^\dagger(\K)\HH(\K)U(\K)\right]_{JJ'}=\epsilon_{\K J}\delta_{J,J'}\,,
\eeq
where the superscript (H) stands for ``Hamiltonian gauge.''  Next we
define
\beq
\label{eq:J-H}
J^{({\rm H})}_{a,JJ'}(\K)=
\begin{cases}
\displaystyle
\frac{i\left\{ U^\dagger(\K) [\partial_a\HH(\K)]   
U(\K)\right\}_{JJ'}}
{\epsilon_{\K J'}-\epsilon_{\K J}} & \text{if $J'\not= J$}\\
0 & \text{if $J'=J$}
\end{cases}
\eeq
and $J_a=UJ^{({\rm H})}_aU^\dagger$. This matrix will only appear in
the combinations $J_a^+=fJ_ag$ and $J_a^-=gJ_af$, where $f$ is the
occupation matrix introduced in \eq{P}, and $g=1-f$.  With these
definitions, the unfolded curvature in the Wannier basis becomes (see
derivation in Appendix~\ref{appendix:curv-unf-wf})
\bea
\label{eq:curv-unf-wf}
\Omega\unf_{ab}(\k_s)&=&
\Re\Tr\left[ Tf(\partial_a\AAA_b-\partial_b\AAA_a)f\right]\nn
&+&2\Im\Tr\left[Tf\AAA_af\AAA_bf\right]\nn
&-&2\Im\Tr\left[T(f\AAA_aJ_b^++J_a^-\AAA_bf+J_a^-J_b^+)\right]\,.\nn
\eea

\equ{curv-unf-wf} is our second important result. It expresses the
unfolded Berry curvature at a point $\k_s$ in the NBZ in terms of the
matrix $T(\k_s,\K)$ given by \eqs{T}{ovlp-wf}, and additional matrices
defined at the folded point $\K$ in the SBZ.  Those other matrices can
be computed from a knowledge of the Hamiltonian and position-operator
matrix elements in the Wannier basis, which are then Fourier
transformed into $\HH(\K)$ and $\AAA_a(\K)$ via
\eqs{HH-K}{AA-K}. Diagonalization of $\HH(\K)$ [\eq{HH-diag}] provides
the energy eigenvalues and rotation matrices used to compute $f(\K)$
and $J_a^\pm(\K)$.\cite{lopez-prb12} Note that the needed derivatives
$\partial_a\HH(\K)$ and $\partial_b\AAA_a(\K)$ are easily obtained by
differentiating \eqs{HH-K}{AA-K}.

It is instructive to consider the trivial unfolding scenario where the
NC and the SC are the same. Then $T$ becomes the identity matrix, the
second term in \eq{curv-unf-wf} vanishes since
$\Im\Tr[\AAA_af\AAA_bf]=0$, and $\boldsymbol{\Omega}\unf(\k)$
correctly reduces to Eq.~(51) of Ref.~\onlinecite{lopez-prb12} for
$\boldsymbol{\Omega}\occ(\k)$.

\section{Computational details}

Plane-wave pseudopotential calculations were carried out for bcc Fe,
bcc Co, and an Fe--Co ordered alloy with the Fe$_3$Al
structure.\cite{schwarz-jpf84} The experimental lattice constant
$a=5.42$~bohr of bcc Fe was used in all cases to facilitate
comparisons, and the magnetization was set along the [001] direction.

The calculations were performed with the {\tt Pwscf} code from the
{\tt Quantum-Espresso} package,\cite{giannozzi-jpcm09} in a
noncollinear spin framework with fully relativistic norm-conserving
pseudopotentials generated from parameters similar to those in
Ref.~\onlinecite{wang-prb06}. An energy cutoff of 120~Ry was used for
the plane-wave expansion of the wavefunctions, and exchange and
correlation effects were treated within the PBE generalized-gradient
approximation.\cite{perdew-prl96}

In the case of bcc Fe and bcc Co, the self-consistent total energy
calculations were done with a $16\times 16\times 16$ Monkhorst-Pack
mesh for the BZ integration, while for the non-self-consistent
calculation a $10\times 10\times 10$ mesh was used, and the 28 lowest
bands were calculated.  In the case of Fe$_3$Co the BZ integration
meshes were $12\times 12\times 12$ and $10\times 10\times 10$ for the
self-consistent and bandstructure calculations respectively, and the
112 lowest bands were calculated. A Fermi smearing of 0.02~Ry was used
in all self-consistent calculations.

For each material, eighteen spinor Wannier functions per atom were
then constructed using {\tt Wannier90}.\cite{mostofi-cpc08}
Atom-center $s$, $p$, and $d$-like trial orbitals were used for the
initial projection step, followed by an iterative procedure to select
an optimal ``disentangled'' subspace,\cite{souza-prb01} using the same
inner and outer energy windows as in Ref.~\onlinecite{wang-prb06}.  At
variance with that work, no minimization of the spread functional was
done to further improve the localization properties of the
``projected'' Wannier functions.\cite{marzari-prb97} This was done to
keep the Wannier functions of Fe$_3$Co as similar as possible to those
of bcc Fe, as required by the Wannier-based unfolding
scheme.\cite{ku-prl10}

In the next section we show results for the energy bands and Berry
curvature of the Fe$_3$Co ordered alloy unfolded onto the NBZ of bcc
Fe. For comparison purposes, we also show the energy bands and Berry
curvatures of pure Fe and of the VCA alloy computed directly in the
NBZ.  Following Ref.~\onlinecite{liu-prb13}, we have implemented the
VCA in the basis of projected Wannier functions, by linearly mixing
the Hamiltonian matrix elements of bcc Fe and bcc Co.  Since the
Wannier interpolation of the Berry curvature also requires the
position-operator matrix elements,\cite{wang-prb06} we modified them
accordingly.

In all cases, with and without unfolding, we plot the Bloch spectral
function instead of the energy bands. To generate the plots we use the
method of Ref.~\onlinecite{berlijn-prl11}, adapted to display the spin
polarization $\langle S_z\rangle$ as a color code. A similar procedure
is used to plot the intersections of the (unfolded) Fermi surface with
a plane in the NBZ. For simplicity, we will continue to use the
expressions ``energy bands'' and ``Fermi surface intersections'' (or
``Fermi lines'') when referring to the figures.

\section{Results}
\label{sec:results}

\begin{figure}
  \centering\includegraphics[width=8.5cm]{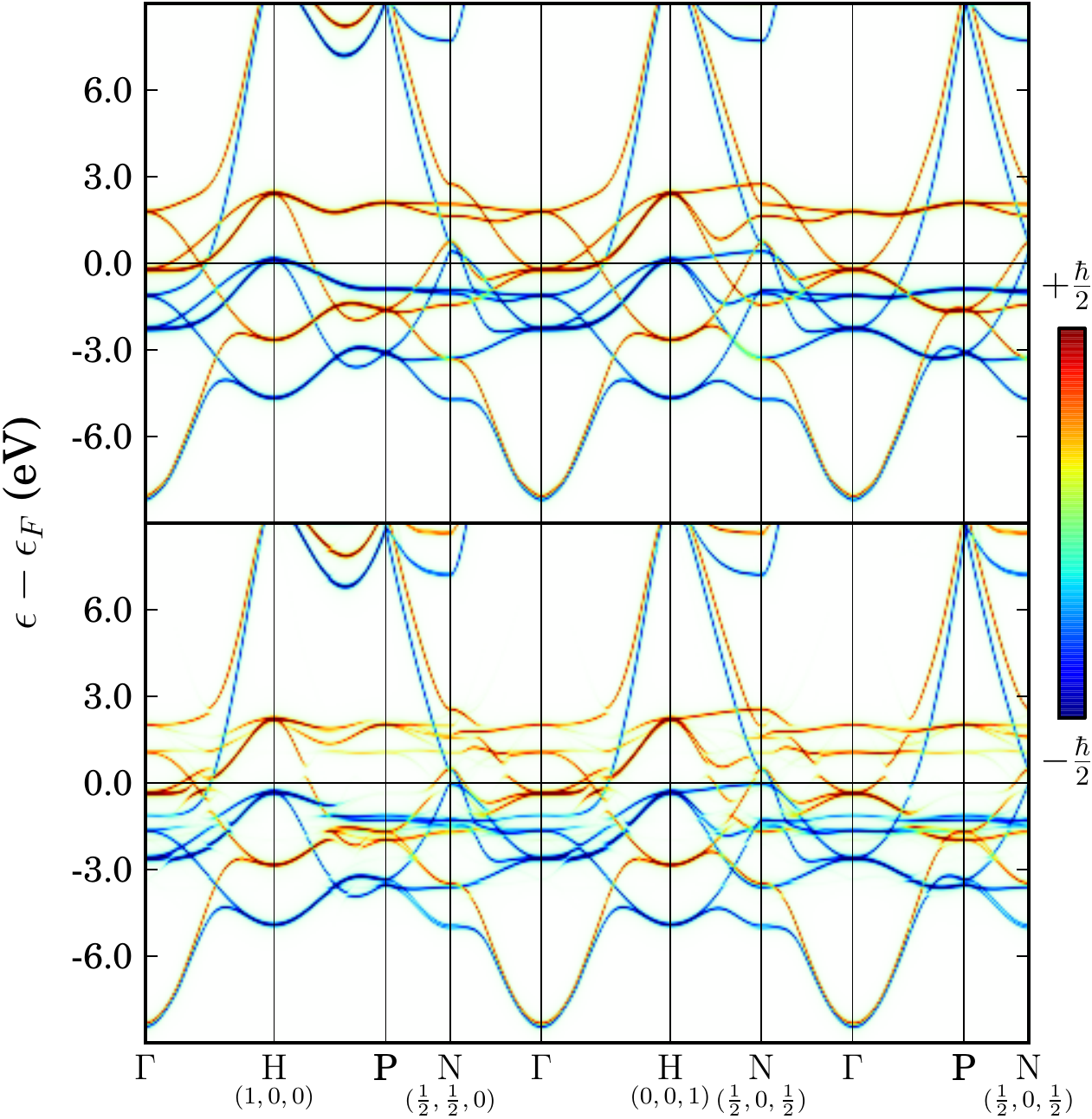}
  \caption{(Color online.) Upper panel: Energy bands of bcc
    Fe. Lower-panel: Energy bands of the fcc Fe$_3$Co alloy unfolded
    onto the Brillouin zone of bcc Fe. Colors indicate the spin
    polarization $\langle S_z\rangle$ of the states.The points
    labelled P all have coordinates
    $(\sfrac{1}{2},\sfrac{1}{2},\sfrac{1}{2})$. The bands of the bcc
    $\langle$Fe$_3$Co$\rangle$ virtual crystal (not shown) are almost
    indistinguishable from those of bcc Fe, except for a shift in the
    Fermi level.}
\label{fig:bands}
\end{figure}

We have selected Fe--Co, a substitutional alloy based on the bcc
structure, as a test case for the Berry curvature unfolding procedure.
We focus on a composition of 25\% Co, using the Fe$_3$Al ordered
structure as a representative configuration.\cite{schwarz-jpf84} This
structure is of the Heusler X$_2$YZ type, with X=Y=Fe and Z=Co (pure
bcc Fe corresponds to X=Y=Z=Fe). The Bravais lattice is fcc with a
four-atom basis, so that the energy bands live in a folded BZ which is
four times smaller than the NBZ of bcc Fe. This makes a direct
comparison with the bandstructure of bcc Fe rather difficult, and
typically only the densities of states are
compared.\cite{schwarz-jpf84}

Information about the $k$-space distribution of the electron states in
the alloy can be recovered by plotting the energy bands unfolded onto
the NBZ [\eq{unfold-bands}]. The result, shown in the lower panel of
Fig.~\ref{fig:bands}, strongly resembles the bands of bcc Fe in the
upper panel.  The influence of the Co atoms is clearly visible in
certain regions of the $(\k,E)$ plane, in the form of ``broken bands''
and ``ghost bands.'' Overall, the effects of alloying are most
pronounced for the narrow $d$ bands crossing the Fermi level.

\begin{figure}
\centering\includegraphics[width=8.15cm]{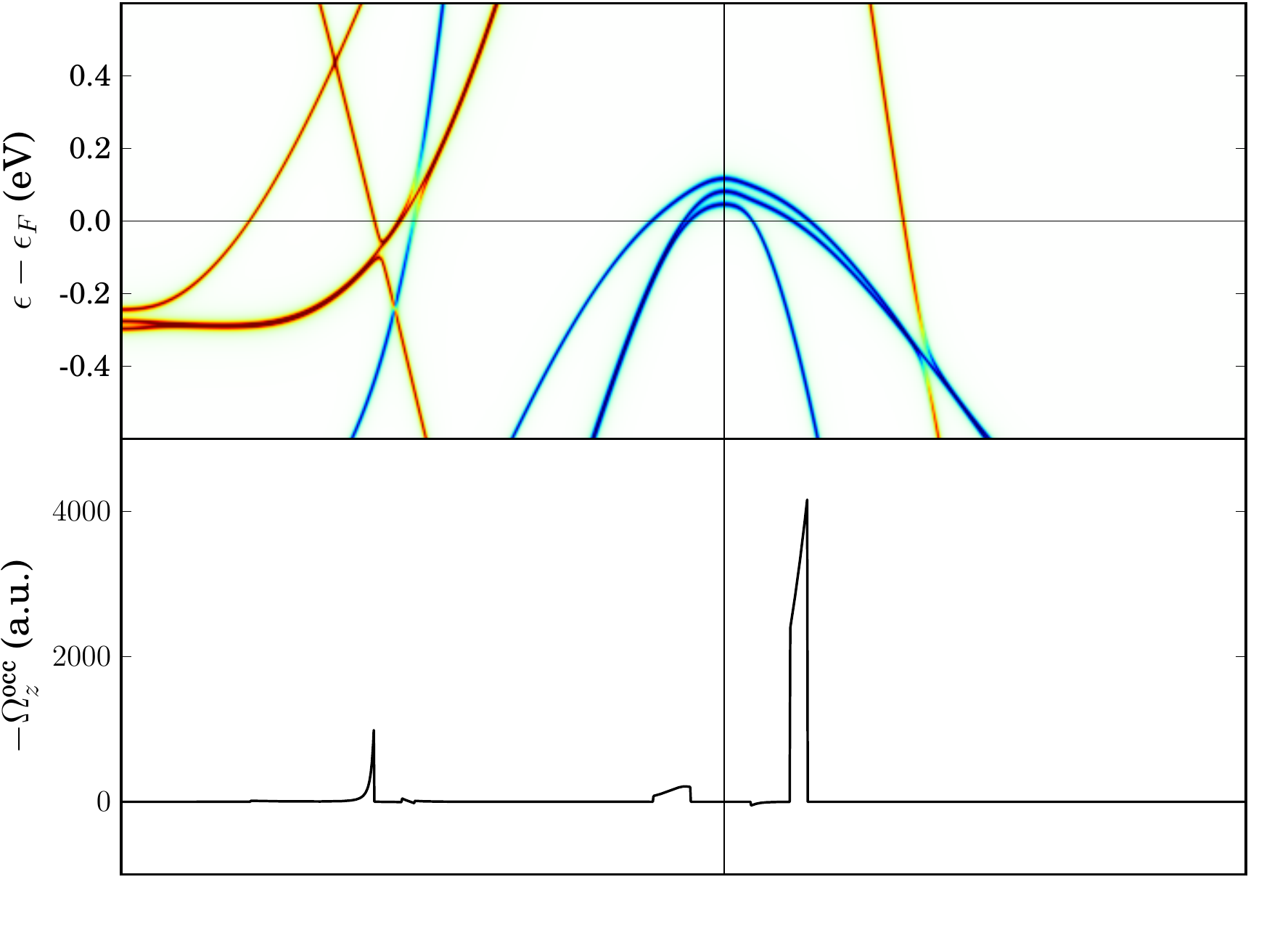}
\centering\includegraphics[width=8.15cm]{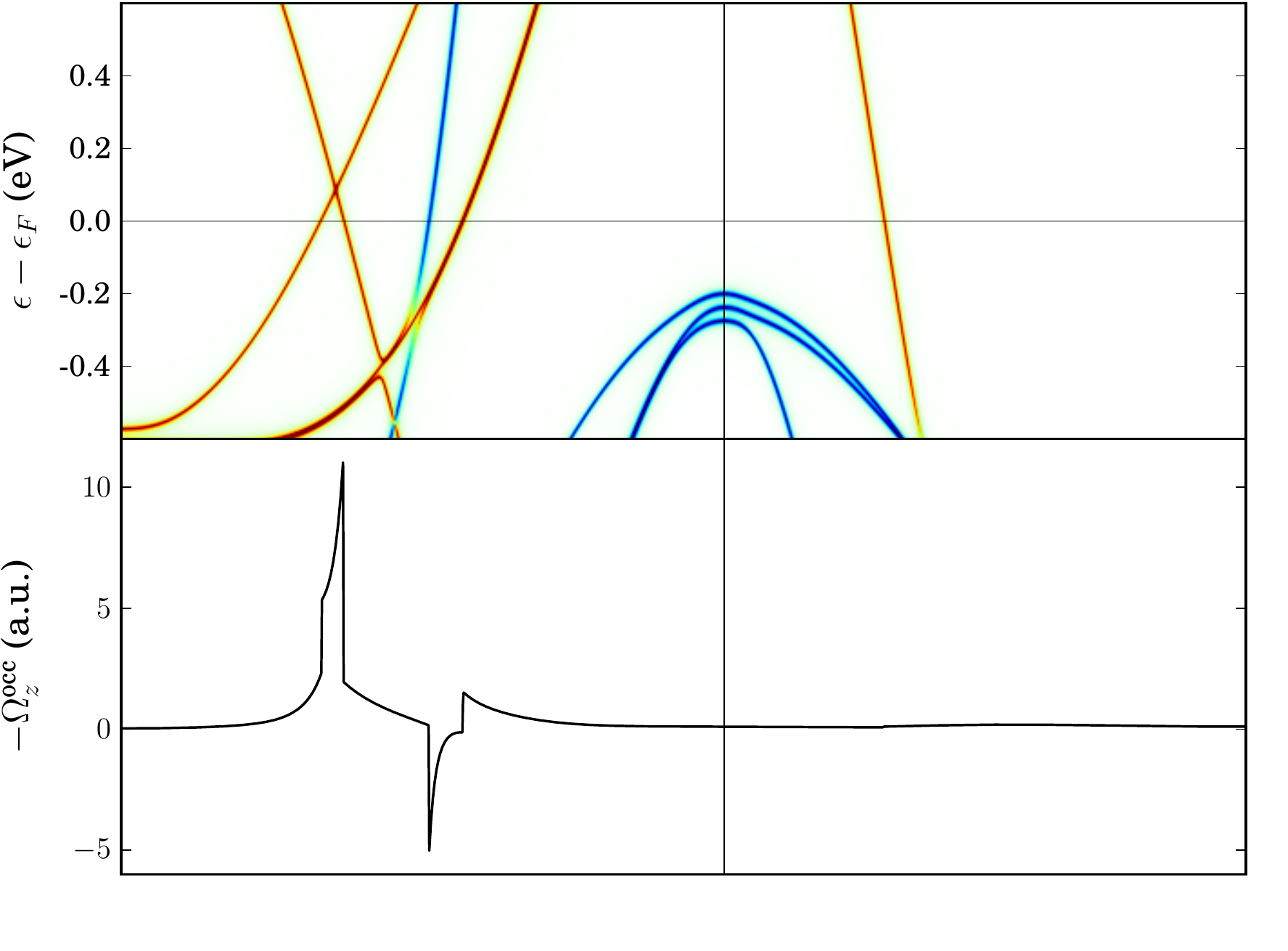}
\centering\includegraphics[width=8.15cm]{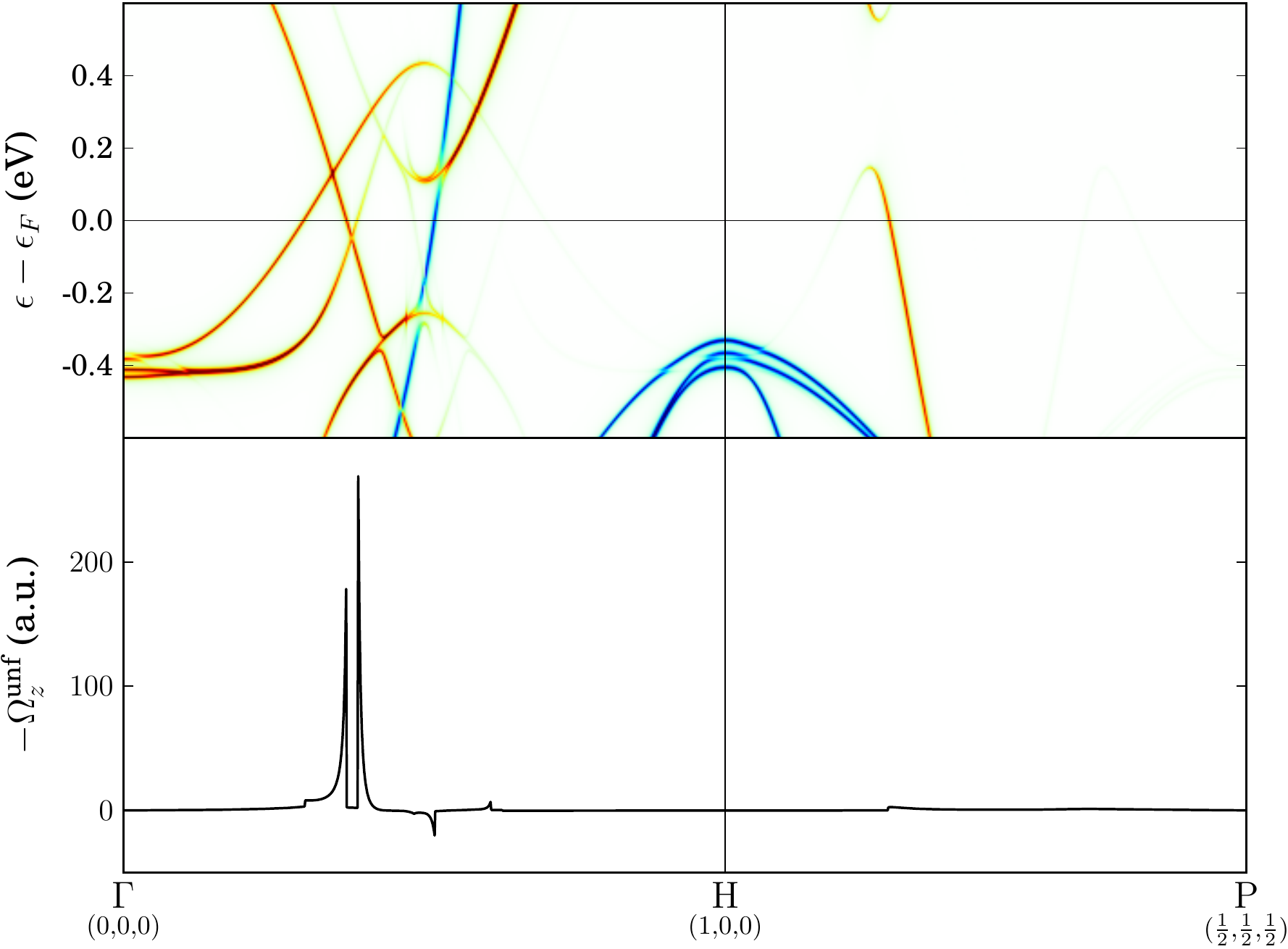}
\caption{(Color online.) Energy bands color-coded by the spin
  polarization $\langle S_z\rangle$ and Berry curvature summed over
  the occupied states, plotted along the path $\Gamma$--H--P.  Upper
  panel: bcc Fe. Middle panel: bcc $\langle$Fe$_3$Co$\rangle$ virtual
  crystal. Lower panel: fcc Fe$_3$Co alloy, using BZ unfolding.}
\label{fig:GHP}
\end{figure}

\begin{figure}
\centering\includegraphics[width=7.15cm]{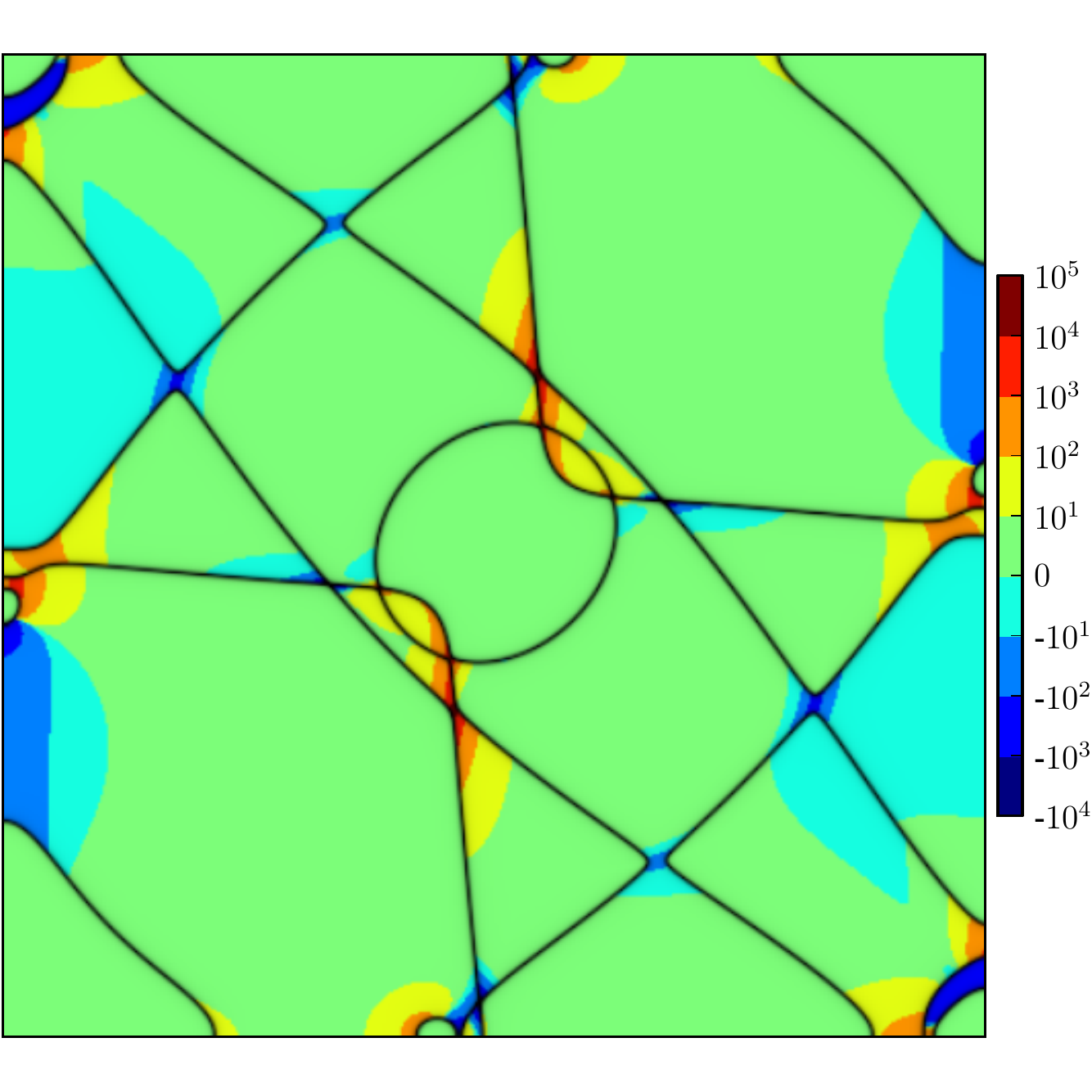}
\centering\includegraphics[width=7.15cm]{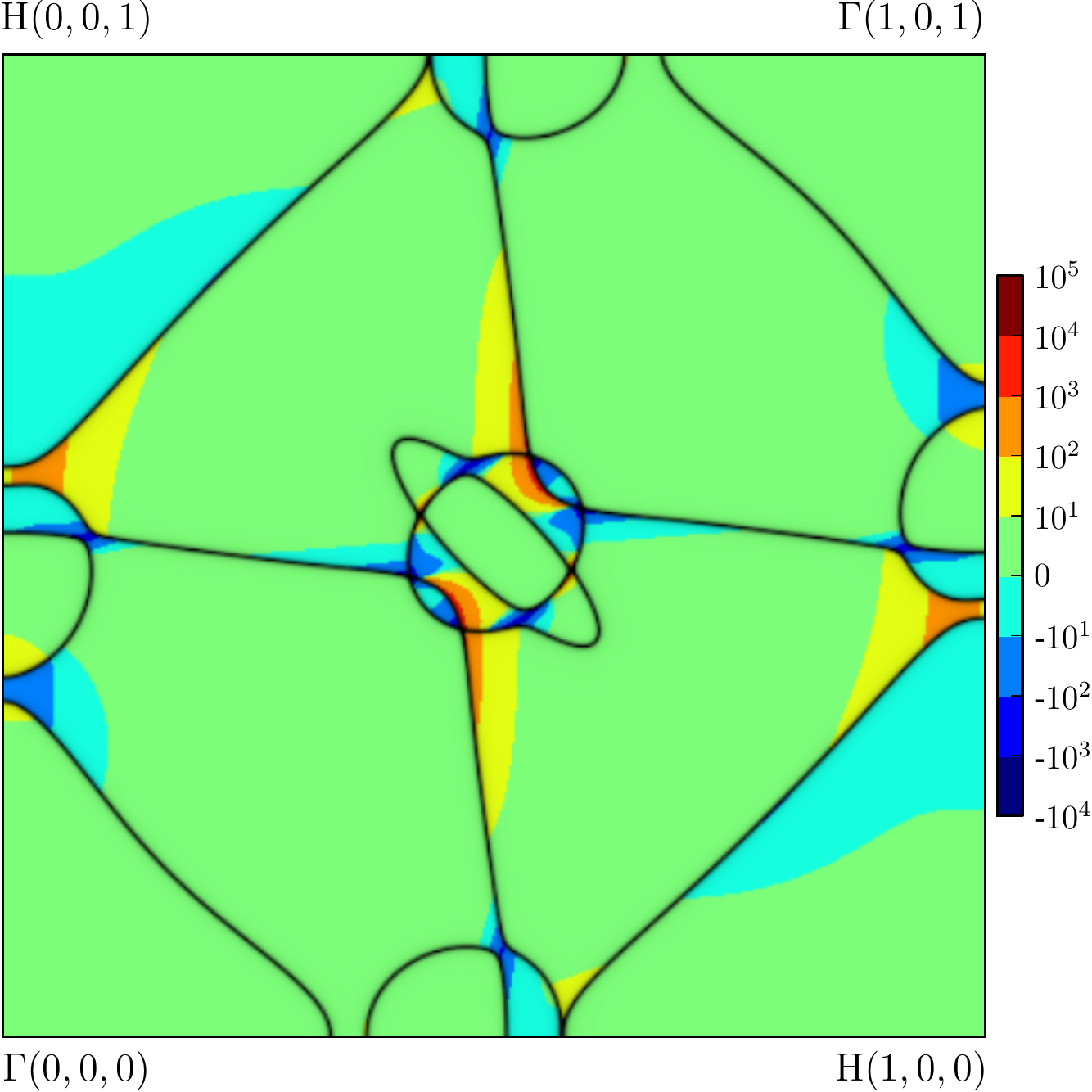}
\centering\includegraphics[width=7.15cm]{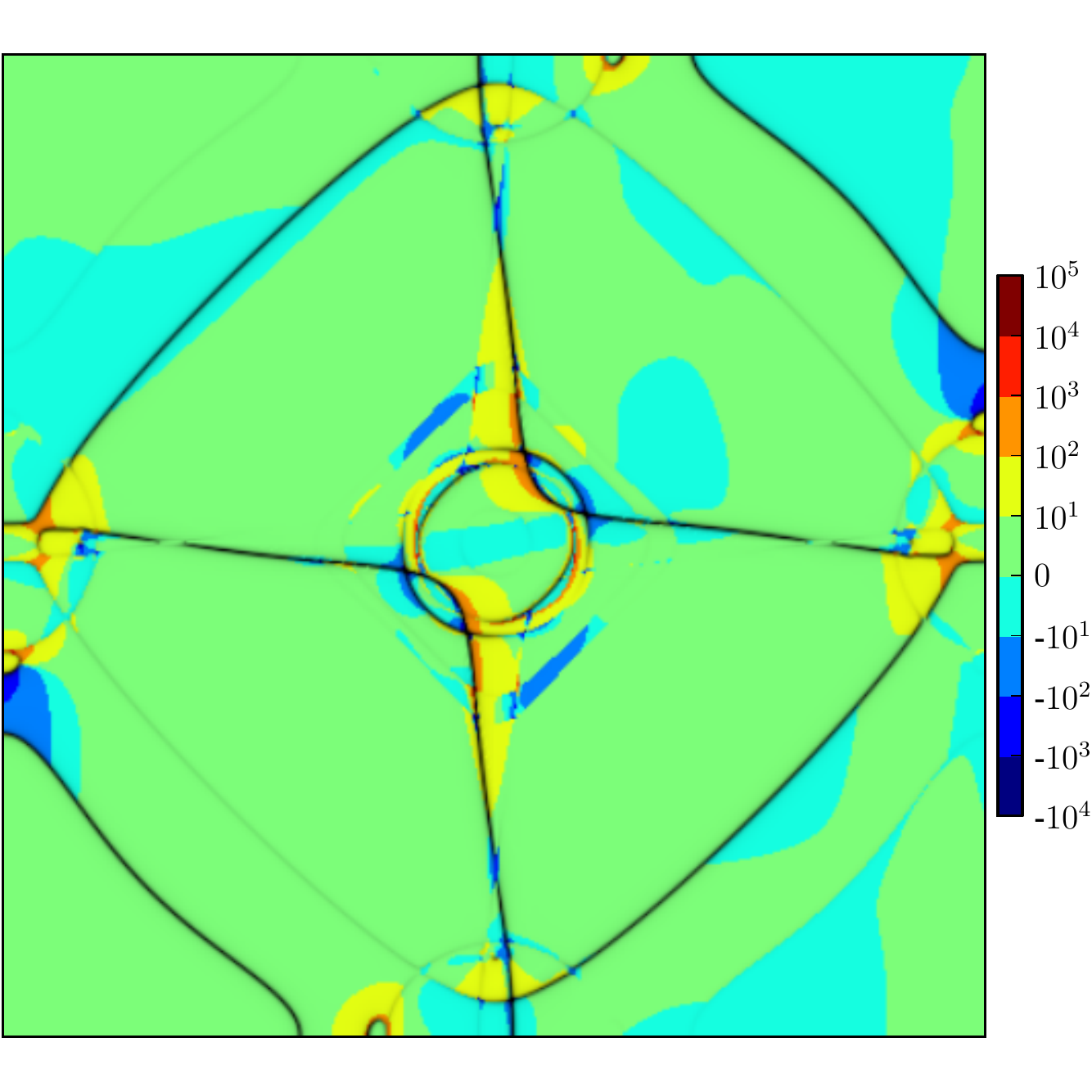}
\caption{(Color online.)  Heatmap of the Berry curvature in the plane
  $k_y=0$, in atomic units (note the log scale). The lines of
  intersection between the Fermi surface and the plane are also shown.
  The upper and middle panels show $-\Omega\occ_z$ for bcc Fe and bcc
  $\langle$Fe$_3$Co$\rangle$ respectively, and the lower panel shows
  $-\Omega\unf_z$ for fcc F$_3$Co.}
\label{fig:heatmaps}
\end{figure}

We now turn to the $k$-space Berry curvature,
\eqs{curv-tot}{curv-kubo-tot}, and begin by recalling its salient
features in crystalline metallic
ferromagnets.\cite{fang-science03,yao-prl04,wang-prb06} In this class
of materials the Berry curvature is induced by the combined effect of
exchange splitting and spin-orbit coupling, which together break
time-reversal symmetry in the orbital
wavefunctions. $\boldsymbol{\Omega}\occ(\k)$ is characterized by
strong, sharp features which are concentrated around the Fermi
surface, in regions where occupied and empty bands come in close
contact and become strongly coupled by spin-orbit. This is illustrated
for bcc Fe in Fig.~\ref{fig:GHP}(a), which displays the energy bands
near the Fermi level and the Berry curvature, along the $\Gamma$--H--P
path. The spiky features rise above a smooth, low-intensity background
which is visible in the heatmap plot of the Berry curvature over the
$k_y=0$ plane, Fig.~\ref{fig:heatmaps}(a).

In order to understand how alloying with Co disturbs the Berry
curvature, we first consider the effects in the virtual-crystal
approximation, that is, for a bcc crystal composed of ``averaged''
$\langle$Fe$_3$Co$\rangle$ atoms.  Since $\k$ remains a good quantum
number in the NBZ, the energy bands and Berry curvature can be
obtained in the usual manner (without unfolding), and are shown in
Fig.~\ref{fig:GHP}(b). The bands are quite similar to those of bcc Fe,
and the main effect of alloying is an upward shift of the Fermi
level. This leads to significant changes in the Berry curvature: for
example, the strong peak along H--P is completely supressed, since the
two majority bands involved are now both occupied. Only some very
low-intensity features remain along $\Gamma$--H (note the difference
in the Berry curvature scales between the panels in
Fig.~\ref{fig:GHP}). 

Comparing the heatmaps in Figs.~\ref{fig:heatmaps}(a,b) we again see
significant differences in the Berry curvature distribution, due to
the shift in the Fermi level across narrow $d$ bands.  In both cases
the Berry curvature is concentrated in regions where there are weak
avoided crossings between two Fermi lines, which can be of
opposite-spin character or of like-spin character.

Missing from the VCA description of the alloy are the effects brought
about by the reduced translational order, which are the main focus of
this work.  Their influence on the bandstructure was revealed by
plotting the unfolded bands of fcc Fe$_3$Co in
Fig.~\ref{fig:bands}. In order to see how the the Berry curvature is
affected, we plot together in Fig.~\ref{fig:GHP}(c) the two unfolded
quantities, energy bands and Berry curvature.

Compared to the VCA results in Fig.~\ref{fig:GHP}(b) the Fermi level
has not moved appreciably, and the bigger changes are in the bands
themselves, especially in the minority states near the Fermi level.
The Berry curvature displays two contiguous strong peaks along
$\Gamma$--H. They are associated with spectral features which have
been greatly modified with respect to the VCA calculation, namely, a
pair of minority bands with a weak avoided crossing just below the
Fermi level. As the upper band rises above $\epsilon_F$ on either side
of the crossing, a Berry-curvature peak suddenly develops and then
quickly drops as the separation between the two bands increases.
Plots along other high-symmetry lines in the NBZ show similar
features.  We conclude that the intuitive ``interband coupling''
interpretation of the Berry curvature based on \eq{curv-kubo-tot}
carries over to the unfolded curvature, now in terms of the unfolded
bands.  Further confirmation of this comes from inspecting the
unfolded Fermi lines and Berry curvature across the $k_y=0$ plane in
Fig.~\ref{fig:heatmaps}(c).  Overall they resemble those of the VCA
crystal, but with some distortions. As before, the Berry curvature is
concentrated in regions where two Fermi lines approach one another.

To conclude we evaluate the AHC of the three systems from
\eqs{ahc}{ahc-SBZ}.  The results were carefully converged with respect
to $k$-point sampling,~\cite{yao-prl04,wang-prb06} using dense uniform
meshes which were adaptively refined around points where the Berry
curvature exceeded a threshold magnitude of 27.98~\AA$^2$.  Uniform
(adaptive) meshes of up to $350\times 350\times 350$ ($13\times
13\times 13$) in the NBZ were used for Fe and
$\langle$Fe$_3$Co$\rangle$.  For Fe$_3$Co the densest uniform
(adaptive) mesh in the SBZ was $250\times 250\times 250$ ($11\times
11\times 11$).  The converged AHC values are 758~S/cm for bcc Fe,
452~S/cm for bcc $\langle$Fe$_3$Co$\rangle$, and 473~S/cm for fcc
Fe$_3$Co. We will comment on these numbers shortly.

\section{Discussion and outlook}
\label{sec:disc}

As illustrated by our calculations, impurities modify the interband
couplings responsible for the intrinsic AHC in perfectly ordered
crystals. In the context of SC calculations it is very natural to
combine the putative intrinsic contribution of \eq{ahc} with those
disorder corrections into a single geometric contribution,
\eq{ahc-SBZ}, which is a gauge-invariant property of the disordered
electronic ground state. In Chern insulators, where the AHE is
quantized for topological reasons (QAHE), the disorder corrections
cancel out upon taking the integral in \eq{ahc-SBZ}. In metals the AHE
is not quantized, and disorder gives a net geometric contribution on
top of the intrinsic one.

In the same way that the intrinsic AHC can be viewed as the dc limit
of the interband conductivity of the pristine
crystal,\cite{nagaosa-rmp10} the geometric AHC corresponds to the dc
limit of the interband conductivity of a SC with disorder, whose
``bands'' are defined in the folded BZ. For disordered systems
possessing a parent ordered structure, the familiar representation in
terms of a Berry curvature in the normal BZ can be partially restored
by means of the unfolded Berry curvature (\ref{eq:curv-unf}), leading
to \eq{ahc-NBZ} which has the same form as \eq{ahc}.

In pristine crystals the geometric AHC reduces to the intrinsic
contribution. It therefore retains the essential features of the
intrinsic theory of the AHE, while at the same time addressing the
main criticism that it originally faced, namely, ``the complete
absence of scattering from disorder in the derived Hall response
contribution.''\cite{nagaosa-rmp10}

Given the reasonably good agreement with experiment which has been
achieved from first principles calculations based on \eq{ahc}, one
should be cautious about introducing modifications.  The calculations
presented in this work are reassuring in that regard: most of the
large change in the calculated AHC between pure bcc Fe and the Fe--Co
alloy is recovered at the VCA ``intrinsic'' level from the
band-filling effect, while ``scattering'' effects from the reduced
translational order in the fcc cell give some corrections, without
dramatically changing the result. The same conclusion can be drawn
from comparing Figs.~\ref{fig:heatmaps}(b,c).

The system we have studied is of course a very crude model for a real
disordered alloy. Calculations using larger SCs with more realistic
descriptions of disorder will be needed to make detailed comparisons
between the (unfolded) Berry curvature of a disordered crystal or
alloy and that of the parent crystal. For example, it seems plausible
that disorder-induced contributions will be smoothened out compared to
the sharp features seen in Figs.~\ref{fig:GHP}(c)
and~\ref{fig:heatmaps}(c). The Wannier-based SC methodology of
Ref.~\onlinecite{berlijn-prl11} seems particularly well-suited for
such studies.

It would be desirable to clarify which scattering contributions are
included in the geometric AHC.  We give a discussion based on the
Kubo-Greenwood (KG) formula for the SC system,\cite{shitade-jpsj12}
written here for $\omega=0$:
\beq
\label{eq:kg}
\sigma_{ab}=\frac{ie^2}{N V}\sum_{\K J J'}\,
\frac{f_{J'}-f_{J}}{\epsilon_{J'}-\epsilon_{J}}
\frac{\me{J}{\hat{v}_a}{J'}\me{J'}{\hat{v}_b}{J}}
     {\epsilon_{J'}-\epsilon_{J}-i\eta}\,,
\eeq
where $V$ is the SC volume and the SBZ is sampled over $N$ points
$\K$.  The full AHC, 
the sum of intrinsic, skew-scattering, and side-jump contributions,
can 
be calculated as the antisymmetric part of \eq{kg}.
Let us recall the role played by the parameter $\eta$: for a finite
volume $V$ the energy levels at fixed $\K$ are discrete, and
absorption becomes impossible at frequencies smaller than the level
spacing. It is for this reason that in SC calculations of the residual
resistivity $\rho_{xx}=1/\sigma_{xx}$ 
one must use a level broadening $\eta(V)$ greater than the mean level
spacing at~$\epsilon_F$.\cite{brown-prl89} Similar considerations
should be relevant for $\sigma_{xy}$, particularly when trying to
recover the skew-scattering contribution, which scales as
$\sigma_{xx}$ and has a similar physical origin.\cite{nagaosa-rmp10}

This analysis suggests that $\sigma^{\rm geom}_{xy}$, which is
obtained from \eq{kg} by taking the $\eta\rightarrow 0^+$ limit at
finite $V$, does not include skew-scattering. Since the longitudinal
conductivity $\sigma_{xx}$ vanishes in that limit, $\sigma^{\rm
  geom}_{xy}$ corresponds to the {\it dissipationless} part of
$\sigma_{xy}$, and this is precisely how the sum of the intrinsic and
side-jump contributions is defined\cite{nagaosa-rmp10} and
measured.\cite{tian-prl09}

Leaving aside matters of definition and interpretation, our
gauge-invariant procedure for unfolding the Berry curvature from SC
calculations seems useful in its own right as an analysis tool
complementary to the unfolding of energy bands.  The $k$-space Berry
curvature induced by interband coherence effects has emerged as a
powerful paradigm to describe the AHE,\cite{nagaosa-rmp10,xiao-rmp10}
and the methods developed in this work seamlessly incorporate disorder
into the picture.

In closing, we mention that the BZ unfolding procedure can be readily
applied to other $k$-space quantities which take the form of traces
over gauge-covariant matrices.  Examples include the occupation
numbers $n(\K)=\Tr f(\K)$, the integrand of the $k$-space orbital
magnetization formula,\cite{ceresoli-prb06,lopez-prb12} and the
quantum metric.\cite{marzari-prb97}

\begin{acknowledgments}

  This work was supported by grants No.~MAT2012-33720 from the
  Ministerio de Econom\'ia y Competitividad (Spain), No.~CIG-303602
  from the European Commission, and by ONR Grant No. N00014-12-1-1041
  (USA).

\end{acknowledgments}

\appendix
\section{Unfolded energy bands}
\label{appendix:bands}

The spectral operator $(E+i\eta-\hat{H})^{-1}$ projected onto the
Bloch space at $\K$ reads, in the SC eigenstate basis,
\beq
\hat{G}_\K(E+i\eta)=\sum_J\,
\frac{\ket{\K J}\bra{\K J}}{E+i\eta-\epsilon_{\K J}}\,.
\eeq
The $\K$-resolved density of states (Bloch spectral function) consists
of sharp peaks in the SBZ, corresponding to the ``folded'' energy
bands:
\bea D_\K(E)&=&-\frac{1}{\pi}\lim_{\eta\rightarrow 0^+}\im\Tr G_\K(E+i\eta)\nn 
&=&\sum_J\,\delta(E-\epsilon_{\K J})\,.  
\eea

Applying the unfolding prescription of \eq{unfold} to the operator
$\hat{\op}=(-1/\pi)\hat{G}_\K(E+i\eta)$ we find
\bea
\label{eq:unfold-bands}
D\unf_{\k_s}(E)&=& 
\lim_{\eta\rightarrow 0^+}\im\tr {\cal O}\uu(\k_i)\nn
&=&\sum_J\,\T_{JJ}(\k_s,\K)\delta(E-\epsilon_{\K J})\,. 
\eea
This is the known expression for the unfolded Bloch spectral
function,\cite{ku-prl10} with
\beq
\T_{JJ}(\k_s,\K)=\frac{1}{|\M|}\sum_n\,\left| \ip{\k_s n}{\K J}\right|^2
\eeq
the spectral weight of $\ket{\K J}$ at $\k_s$.  (The factor of
$1/|\M|$ on the right-hand-side is absent when adopting the
normalization convention of Ref.~\onlinecite{ku-prl10}.)

\section{Unfolding sum rule}
\label{appendix:sum-rule}

As mentioned in Sec.~\ref{sec:basic}, the unfolding weights satisfy
$\sum_s\,\T_{NN}(\k_s,\K)=1$.  To find the resulting sum rule for
$\op\unf(\k)$, evaluate \eq{unfold} in a basis where either
$\T(\k_s,\K)$ or $\op(\K)$ is diagonal, and sum over $\k_s$:
\beq
\sum_{s=1}^{|\M|}\,\op\unf(\k_s)=
\sum_{s=1}^{|\M|}\sum_N\,\T_{NN}(\k_s,\K)\op_{NN}(\K)
=\Tr\op(\K)\,.
\eeq
(This corresponds to \eq{curv-sum-rule} for the Berry curvature.)  Now
sum over a uniform grid in the SBZ, replace $\sum_\K^{\rm SBZ}\sum_s$
on the left-hand side with $\sum_\k^{\rm NBZ}$, and take the continuum
limit to find
\beq
\label{eq:sbz-nbz}
\int_{\rm NBZ}d^3k\,\op\unf(\k)=
\int_{\rm SBZ}d^3K\,\Tr\op(\K)\,,
\eeq
which corresponds to \eqs{ahc-SBZ}{ahc-NBZ}.

\section{Derivation of \eq{curv-unf-wf}}
\label{appendix:curv-unf-wf}

In this Appendix we derive \eq{curv-unf-wf} for
$\Omega\unf_{ab}(\k_s)$ starting from \eq{curv-unf}.  Folowing
Ref.~\onlinecite{lopez-prb12}, we adopt a notation where matrix
objects written with a double staff, such as $\AAA_{NM}(\K)=i\ip{u_{\K
    N}}{\partial_a u_{\K M}}$ in \eq{AA-K}, are defined over the space
spanned by the Wannier functions, which for metals typically contains
some low-lying empty states in addition to all the occupied
states.\cite{souza-prb01} Instead, objects with a single staff such as
$A_{NM}(\K)$ in \eq{nonabelian-ins} are defined over the occupied
subspace. So, for example, we define (dropping $\K$ everywhere)
$\hat{\PP}=\sum_N\,\ket{u_N}\bra{u_N}$,
$\hat{\QQ}=\hat{\mathbbm{1}}-\hat{\PP}$, and $\FF_{ab,NM}=
i\me{\partial_a u_N}{\hat{\QQ}}{\partial_b u_M}$ as counterparts to
$\hat{P}$, $\hat{Q}$, and $F_{ab,NM}$ in \eqs{F-ab}{P}.

We further condense our notation by dropping indices $N$, $M$, e.g.,
$\PP=\ket{u}\bra{u}$ and $\hat{P}=\ket{u}f\bra{u}$. We will use the
relations\cite{lopez-prb12}
\beq
\label{eq:rel-a}
(\partial_a\hat{P})\hat{Q}=\ket{u}f\bra{\partial_a u}\hat{\QQ}+
i\ket{u}f(\AAA_a+J_a)g\bra{u}
\eeq
and (compare with \eqs{nonabelian-ins}{nonabelian})
\beq
\label{eq:rel-b}
i\FF_{ab}-i\FF_{ba}=\partial_a\AAA_b-\partial_b\AAA_a-i[\AAA_a,\AAA_b]\,.
\eeq 
Expanding \eq{F-ab} with the help of \eq{rel-a} we find
\beq
\label{eq:F-ab-wf}
F_{ab}=f\FF_{ab}f+f\AAA_ag\AAA_bf+J^-_a\AAA_bf+f\AAA_aJ^+_b+J^-_aJ^+_b\,.
\eeq
Writing \eq{curv-unf} as $\Omega\unf_{ab}= i\Tr
\left\{T\left[F_{ab}-F_{ba}\right]\right\}$, inserting \eq{F-ab-wf}
and then using \eq{rel-b}, we arrive at \eq{curv-unf-wf}.

%

\end{document}